\documentclass{article}

\PassOptionsToPackage{numbers, compress}{natbib}

\usepackage[preprint]{neurips_2025}
\usepackage{float}
\usepackage{pifont}
\usepackage{footnote}
\usepackage{enumitem}
\usepackage{bm}
\usepackage{arydshln}
\usepackage{booktabs}
\usepackage{multicol}
\usepackage{multirow}
\usepackage{color}
\usepackage{xcolor}     
\usepackage{colortbl}
\usepackage{soul}
\usepackage{bbding}
\usepackage{makecell}
\usepackage{mathtools}
\usepackage{imakeidx}
\usepackage{setspace}
\usepackage{threeparttable}
\makeindex
\usepackage{arydshln}
\usepackage{lipsum}
\usepackage{natbib}
\usepackage[toc]{multitoc}
\usepackage[edges]{forest}
\usepackage[normalem]{ulem}

\usepackage{bbding}
\usepackage[most]{tcolorbox}

\usepackage{algorithm}
\usepackage{algorithmic}

\usepackage{minitoc}
\usepackage[toc,page,header]{appendix}

\definecolor{orchid}{rgb}{0.85, 0.44, 0.84}
\definecolor{rubinred}{rgb}{0.82, 0.0, 0.28}
\definecolor{flagship}{rgb}{0.93, 0.06, 0.41}
\definecolor{radiologist}{rgb}{0.50, 0.50, 1}

\newcommand{\dataset}{{\fontfamily{ppl}\selectfont
PanTS}}

\newcolumntype{P}[1]{>{\centering\arraybackslash}p{#1}}
\newlength\savewidth
  
\usepackage{array,graphicx,xcolor,booktabs,threeparttable}
\newcommand{\yes}{\textcolor{rubinred}{$\checkmark$}}
\newcommand{\no}{\textcolor{gray}{$\times$}}
\newcolumntype{C}{>{\centering\arraybackslash}p{0.018\linewidth}}
\newcolumntype{W}{>{\centering\arraybackslash}p{0.06\linewidth}}

\usepackage[utf8]{inputenc} 
\usepackage[T1]{fontenc}    
\usepackage{url}            
\usepackage{booktabs}       
\usepackage{amsfonts}       
\usepackage{nicefrac}       
\usepackage{microtype}      
\usepackage{xcolor}         
\usepackage{graphicx}
\usepackage{multirow}
\usepackage{pifont}
\usepackage{amssymb}

\usepackage{makecell}
\usepackage{threeparttable}
\usepackage{array}

\definecolor{citecolor}{HTML}{0071BC}
\definecolor{linkcolor}{HTML}{ED1C24}
\usepackage[colorlinks,
            anchorcolor=red,
            citecolor=citecolor, 
            linkcolor=linkcolor,
            ]{hyperref}
            
\newcolumntype{P}[1]{>{\centering\arraybackslash}p{#1}}

\newcommand{\etal}{\textit{et al. }}

\title{\dataset: The Pancreatic Tumor Segmentation Dataset}

\author{
\bf Wenxuan Li\textsuperscript{1}\thanks{Equal contribution.} \quad
\bf Xinze Zhou\textsuperscript{1}\footnotemark[1] \quad
\bf Qi Chen\textsuperscript{1}\footnotemark[1] \quad
\bf Tianyu Lin\textsuperscript{1} \quad
\bf Pedro R. A. S. Bassi\textsuperscript{1,2,3} \\
\bf Szymon Płotka\textsuperscript{4} \quad
\bf Jarosław B. Ćwikła\textsuperscript{5,6} \quad
\bf Xiaoxi Chen\textsuperscript{7} \quad
\bf Chen Ye\textsuperscript{8} \quad
\bf Zheren Zhu\textsuperscript{9,10} \\
\bf Kai Ding\textsuperscript{11} \quad
\bf Heng Li\textsuperscript{11} \quad
\bf Kang Wang\textsuperscript{9} \quad
\bf Yang Yang\textsuperscript{9} \quad
\bf Yucheng Tang\textsuperscript{12} \quad
\bf Daguang Xu\textsuperscript{12} \\
\bf Alan L. Yuille\textsuperscript{1} \quad
\bf Zongwei Zhou\textsuperscript{1}\thanks{Correspondence to: Zongwei Zhou (\href{mailto:zzhou82@jh.edu}{\textsc{zzhou82@jh.edu}})} \\[2.5mm]
\textsuperscript{1}Department of Computer Science, Johns Hopkins University \\
\textsuperscript{2}Department of Pharmacy and Biotechnology, University of Bologna \\
\textsuperscript{3}Center for Biomolecular Nanotechnologies, Istituto Italiano di Tecnologia \\
\textsuperscript{4}Faculty of Mathematics and Computer Science, Jagiellonian University \\
\textsuperscript{5}Department of Cardiology and Internal Medicine, University of Warmia and Mazury \\
\textsuperscript{6}Diagnostic and Treatment Center Gammed \\
\textsuperscript{7}Department of Bioengineering, University of Illinois Urbana-Champaign \\
\textsuperscript{8}Department of General Surgery, Peking University Third Hospital \\
\textsuperscript{9}Department of Radiology \& Biomedical Imaging, University of California, San Francisco \\
\textsuperscript{10}Department of Bioengineering, University of California, Berkeley \\
\textsuperscript{11}Department of Radiation Oncology, Johns Hopkins School of Medicine \\
\textsuperscript{12}NVIDIA \\[2mm]
{\small \texttt{Code, Models \& Data:} \href{https://github.com/MrGiovanni/PanTS}{\texttt{https://github.com/MrGiovanni/PanTS}}}
}

\begin{document}

\maketitle

\doparttoc 
\faketableofcontents 

\begin{abstract}

\dataset\ is a large-scale, multi-institutional dataset curated to advance research in pancreatic CT analysis. It contains 36,390 CT scans from 145 medical centers, with expert-validated, voxel-wise annotations of over 993,000 anatomical structures, covering pancreatic tumors, pancreas head, body, and tail, and 24 surrounding anatomical structures such as vascular/skeletal structures and abdominal/thoracic organs. Each scan includes metadata such as patient age, sex, diagnosis, contrast phase, in-plane spacing, slice thickness, etc. AI models trained on \dataset\ achieve significantly better performance in pancreatic tumor detection, localization, and segmentation than those trained on existing public datasets. Our analysis indicates that these gains are directly attributable to the 16$\times$ larger-scale tumor annotations and indirectly supported by the 24 additional surrounding anatomical structures. As the largest and most comprehensive resource of its kind, \dataset\ offers a new benchmark for developing and evaluating AI models in pancreatic CT analysis.

\end{abstract}

\section{Introduction}
\label{sect:intro}

\begin{figure}[t]
\centering  
    \includegraphics[width=1\linewidth]{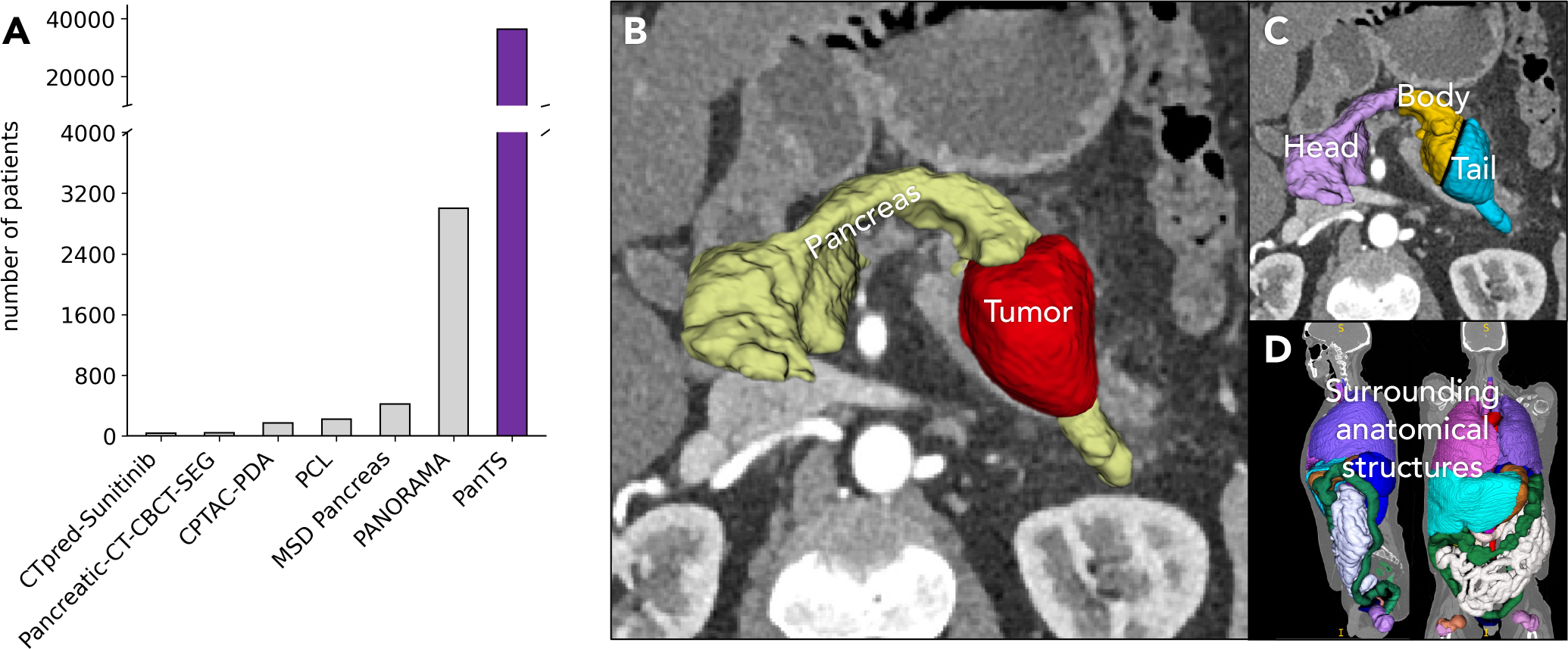}
    \caption{\textbf{Dataset characteristics and visualization.}
    \textbf{A.} \dataset\ comprises 36,390 CT scans collected from 145 medical centers, paired with expert-validated voxel-wise annotations, 16$\times$ larger than the biggest public dataset (i.e., PANORAMA \cite{alves2024panorama}) to date. \textbf{B--C.} The dataset includes detailed annotations for pancreatic tumors, pancreas, and its head, body, and tail, enabling spatially aware tumor localization. \textbf{D.} Twenty-four surrounding anatomical structures are voxel-wise annotated to provide rich spatial context, including key vessels, ducts, and organs critical for tumor detection, resectability assessment, and radiotherapy planning.
    }
    \label{fig:dataset_characteristics}
\end{figure}

Pancreatic cancer is the third leading cause of cancer-related death in the U.S. in both men and women combined \cite{siegel2025cancer, siegel2024cancer,xia2022felix}. Yet despite its clinical importance, early detection remains a major challenge due to the absence of disease-specific symptoms and the incidental nature of abdominal imaging \cite{mcguigan2018pancreatic}. Consequently, 80--85\% of pancreatic tumors are diagnosed at advanced stages, when treatment options are limited and prognosis is poor \cite{zhang2018challenges}. In contrast, early-stage tumors are associated with markedly better outcomes, emphasizing the urgent need for earlier identification \cite{zhao2020pancreatic}.

Computed tomography (CT), especially with contrast enhancement, is the primary modality for evaluating pancreatic abnormalities \cite{chu2017diagnosis}. Retrospective studies have shown that early radiographic signs---such as ductal dilation or focal atrophy---can appear months before clinical diagnosis, but often go undetected \cite{hoogenboom2021pancreatic,chung2022clinical,konno2023retrospective}. However, these indicators are frequently missed in clinical practice, particularly when scans are acquired for unrelated reasons \cite{singh2020computerized,toshima2021ct}. Pancreatic tumors in CT scans are highly heterogeneous in shape, size, location, and radiologic appearance \cite{mukherjee2022radiomics}.

Recent advances in AI have shown promise in automating the detection and localization of pancreatic tumors in CT scans \cite{cao2023large,li2023early,li2024text}. However, most publicly available models are trained on small, homogeneous datasets and fail to generalize to diverse clinical settings. This shortcoming reflects a fundamental data limitation: the pancreas is a small, anatomically intricate organ embedded among critical vessels, ducts, and adjacent structures, making comprehensive annotation and assessment particularly challenging \cite{huang2022artificial,li2025scalemai,li2024abdomenatlas}. Accurate analysis of pancreatic tumors depends not only on identifying the tumor itself but also on understanding its anatomical context.

To address this limitation, we present the \ul{Pan}creatic \ul{T}umor \ul{S}egmentation Dataset (\dataset)---the largest and most comprehensive dataset to date for pancreatic CT analysis\footnote{\dataset\ is not intended for direct clinical decision-making or real-time diagnosis.}. \dataset\ comprises 36,390 CT scans from 145 medical centers. Each scan is paired with metadata, including patient age, sex, contrast phase, diagnosis, in-plane spacing, and slice thickness. Importantly, \dataset\ includes over 993,000 expert-validated voxel-wise annotations (examples in Figure~\ref{fig:dataset_characteristics}), covering:

\begin{itemize}

    \item Pancreatic tumors along with pancreas head, body, and tail, to enable tumor detection, localization, and segmentation. We find that increasing the number of annotated tumors \textit{directly} improves AI performance on out-of-distribution datasets (Figure \ref{fig:justification_large_scale}). To this end, a team of 23 radiologists have produced voxel-wise tumor annotations in each CT scan to support effective AI training at scale.
    
    \item Twenty-four surrounding anatomical structures (e.g., superior mesenteric artery, bile ducts; full list in \S\ref{sec:dataset}) are annotated to enable comprehensive tumor analysis. Joint training on tumors and nearby structures \textit{indirectly} enhances AI performance by reducing false positives and providing rich anatomical context (Figure~\ref{fig:justification_more_structures}). Feature analysis reveals that models trained with both tumor and anatomical structure labels learn more discriminative and separable representations, allowing for more precise tumor detection and segmentation.
    
\end{itemize}

With its large scale, diversity, and anatomical detail, \dataset\ sets a new benchmark for AI development in pancreatic CT analysis. It includes 9,901 publicly available training scans (non-commercial license) and 26,489 test scans reserved for third-party evaluation. This setup follows best practices in medical AI benchmarking \cite{ma2024automatic,bassi2024touchstone,antonelli2022medical,li2024well}, ensuring fair and reproducible comparisons. We also release a strong baseline model, nnU-Net, alongside the dataset. This baseline model ranked Top-1 in the official \href{https://decathlon-10.grand-challenge.org/evaluation/challenge/leaderboard/}{Medical Segmentation Decathlon (MSD) Leaderboard}.

\section{Related Datasets \& Our Contribution}\label{sec:related_work}

\subsection{Pancreas and Other Organ Datasets} 

Several public datasets have advanced multi-organ segmentation in abdominal CT, including BTCV \cite{landman2015miccai} (50 CTs, 13 classes, 1 center), CHAOS \cite{kavur2021chaos} (40 CTs, 4 class, 1 center), AMOS22 \cite{ji2022amos} (500 CTs, 15 classes, 2 centers), WORD \cite{luo2022word} (150 CTs, 16 classes, 1 center), and AbdomenCT-1K \cite{ma2020abdomenct} (1,112 CTs, 4 classes, 12 centers). These datasets typically target general abdominal structures or liver segmentation, with limited diversity in institution count ($\leq$12 centers) and relatively modest dataset sizes. TotalSegmentator \cite{wasserthal2023totalsegmentator} is one of the most ambitious efforts to date, offering 1,228 CT scans across 117 classes from a single source. However, its focus remains on broad anatomic structure segmentation and lacks dedicated design for oncologic applications.

\textbf{\textit{Limitation:}} While these datasets are useful for general anatomical segmentation, they are not specifically designed for pancreatic tumor analysis. None of them provides voxel-wise annotations of important pancreatic substructures, such as the head, body, and tail of the pancreas, the superior mesenteric artery, pancreatic duct, common bile duct, celiac artery, and duodenum. These annotations are essential for surgical decision-making, tumor staging, and accurate assessment of tumor invasion and resectability. Reference organs such as the liver, spleen, kidneys, adrenal glands, aorta, and postcava are either inconsistently labeled or absent \cite{liu2023clip,liu2024universal,kang2023label,zhu2022covid,zhang2023continual,tang2024efficient}. Furthermore, distal anatomical landmarks, including the lungs, femurs, bladder, and prostate, which are important for spatial orientation and radiotherapy planning, are rarely included.

\textbf{\textit{Our Contribution:}} \dataset\ addresses these limitations by offering voxel-wise annotations for 27 clinically meaningful structures selected specifically to support pancreatic tumor analysis. These include voxel-wise annotations of the pancreas head, body, and tail, and 24 surrounding anatomical structures crucial for spatial reasoning, proximity assessment, and downstream clinical workflows such as radiotherapy planning and vessel invasion analysis. With 36,390 CT scans from 145 global medical centers, \dataset\ is not only the largest organ segmentation dataset available, but also the most diverse—offering over 3$\times$ more institutional representation and over 7$\times$ more data than leading datasets like AbdomenCT-1K \cite{ma2020abdomenct} or AMOS22 \cite{ji2022amos}.

\subsection{Pancreatic and Other Tumor Datasets} 

Tumor segmentation datasets have historically focused on more common cancers and organs. For instance, liver tumors are supported by datasets like LiTS \cite{bilic2019liver} (201 CTs, 7 centers), HCC-TACE-Seg \cite{HCC-TACE-seg-Moawad2021} (105 CTs), and MSD Liver \cite{antonelli2022medical} (201 CTs); colorectal tumors by Stagell-Colorectal-CT \cite{StageII-Colorectal-CT-Tong2022} (230 CTs); kidney tumors by TCGA-KIRC \cite{TCGA-KIRC-Akin2016} (267 CTs) and KiTS23 \cite{heller2019kits19} (599 CTs); and lung tumors by MSD Lung \cite{antonelli2022medical} (96 CTs). Large-scale efforts such as FLARE23 \cite{FLARE23-ma2024automaticorganpancancersegmentation} (4,500 CTs, 14 classes, more than 50 centers) and autoPET \cite{autoPET-Aerts2022} (1,214 CTs, 1 class) target pan-cancer analysis but lack pancreas-specific detail or annotations of relevant anatomical structures.

\textbf{\textit{Limitation:}} Pancreatic tumor datasets, in comparison, remain scarce and small in scale \cite{bassi2025radgpt,bassi2024label,chou2024embracing}. NIH Pancreas-CT \cite{PCL-diagnostics11050901} (82 CTs), Pancreatic-CT-CBCT-SEG \cite{hong2021breath} (40 CTs), and CPred-Sunitinib-panNET \cite{chen2023special} (38 CTs) are all limited to single centers and focus on narrow tumor types or clinical scenarios. PANORAMA \cite{alves2024panorama} (2,238 CTs, 6 classes, 7 centers) is a major step forward, offering voxel-wise annotations for pancreatic ductal adenocarcinoma (PDAC) and associated structures such as ducts and vessels. However, it does not provide annotations for other types of pancreatic tumors, which causes issue in evaluation as discussed in \S\ref{sec:scaling}.

\textbf{\textit{Our Contribution:}} \dataset\ is the largest and most comprehensive publicly available dataset for pancreatic tumor segmentation, offering over 16$\times$ more annotated CT scans than PANORAMA and spanning over 20$\times$ more medical centers. In addition to voxel-wise annotations of pancreatic tumors, \dataset\ provides segmentation of the pancreas head, body, and tail, enabling precise tumor localization and region-aware staging. The dataset supports a full pipeline of clinically relevant tasks—tumor detection, segmentation, staging, resectability assessment, and surgical planning—by also including 24 surrounding anatomical structures critical for evaluating tumor involvement of vessels and adjacent organs. No existing dataset provides this combination of scale, diversity, and task-aligned anatomical detail.

\section{\dataset: The Pancreatic Tumor Segmentation Dataset}\label{sec:dataset}

\dataset\ comprises 36,390 CT scans with precise per-voxel annotations of pancreatic tumors, pancreas head, body, and tail, along with 24 surrounding structures (i.e., pancreas, superior mesenteric artery, pancreatic duct, celiac artery, common bile duct, veins, aorta, gall bladder, left and right kidneys, liver, postcava, spleen, stomach, left and right adrenal glands, bladder, colon, duodenum, left and right femurs, left and right lungs, and prostate). Sourced from 145 centers, this dataset includes imaging metadata such as patient sex, age, contrast phase, diagnosis, spacing, and scanner details.

We split the \dataset\ into a training set of 9,901 cases (27\%) and a test set of 26,489 cases (73\%). Both sets consist of abdominal CT scans. Detailed dataset characteristics are summarized in Table~\ref{tab:pants_stats}. The data and annotation are licensed as CC BY-NC-SA. We have released the training set to \href{https://huggingface.co/datasets/AbdomenAtlas/PanTSMini}{The \dataset\ Huggingface Website}, and the test set is preserved for thirty-party evaluation. 



\subsection{Dataset Diversity}

The \dataset\ dataset comprises a broad spectrum of pancreatic tumor types, including pancreatic ductal adenocarcinoma, pancreatic neuroendocrine tumors (PNETs), pancreatic cystic neoplasms, and cystic non-neoplastic lesions. These entities exhibit heterogeneous imaging characteristics in terms of size, morphology, attenuation, and texture. The CT scans are abdominal images obtained using varying contrast phases, scanner models, and imaging protocols. The dataset also contains real-world imaging artifacts, such as metal-induced streaks, contributing to substantial variability in spatial resolution and image quality. The number of tumors per case ranges from 1 to 6, and tumor sizes range from 4 mm to 68 mm in diameter. The test set contains a higher frequency of tumor occurrences than the training set. The average Hounsfield Unit (HU) value of tumors is 60.8 in the training set and 67.7 in the test set. Dataset statistics are summarized in Table~\ref{tab:pants_stats}. The training and test sets follow a 2:1 split and originate from entirely different centers. Therefore, \dataset\ allows thorough evaluation of AI generalization to unseen institutions.

\begin{table}[h]
    \centering
    \caption{\textbf{Characteristics of the \dataset\ dataset.} 
    The \dataset\ training and test sets differ significantly across most clinical and imaging variables, including age, sex distribution, image resolution, and contrast phases. $p$-values were computed with the Mann–Whitney $U$ test. Notably, the test set contains a similar proportion of tumor cases but includes more non-contrast scans, making it a more challenging and realistic out-of-distribution benchmark. Tumor burden and pancreas size also vary between sets, reinforcing the need for robust generalization in model evaluation. These differences justify our dataset split design for assessing model performance under distributional shifts.
    }
    \label{tab:pants_stats}
    \scriptsize
    \begin{tabular}{p{0.001\linewidth}p{0.2\linewidth}P{0.26\linewidth}P{0.26\linewidth}P{0.11\linewidth}}
        \toprule
        \multicolumn{2}{l}{Variable} & Training set ($n$ = 9,901) & Test set ($n$ = 26,489) & $p$-value \\
        \midrule
        \multicolumn{2}{l}{Age, mean (SD)} & 50.8 (20.4) & 58.8 (17.0) & $1.2\times 10^{-172}$\\
        \multicolumn{3}{l}{Sex} & &  $2.1\times 10^{-182}$\\
         & Female, no. (\%) & 2,976 (30.1) & 13,091 (49.4) \\
         & Male, no. (\%) & 4,460 (45.0) & 11,714 (44.2)\\
         & Unknown, no. (\%) & 2,465 (24.9) & 1,684 (6.4)\\
        \multicolumn{2}{l}{In-plane spacing, mm (IQR)}  & 0.81 (0.73, 0.92) & 0.75 (0.70, 0.83) & $1.1\times 10^{-300}$\\
        \multicolumn{2}{l}{Slice thickness, mm (IQR)}   &  2.50 (0.87, 3.00) & 1.25 (1.25, 2.50) & $1.0\times 10^{-63}$\\
        \multicolumn{2}{l}{Contrast phase} &  & &$4.3\times 10^{-45}$\\
         & Non-contrast, no. (\%) & 781 (7.9) & 3,921 (14.8)\\
         & Portal venous, no. (\%) & 6,422 (64.9) & 20,295 (76.6)\\
         & Arterial, no. (\%) & 2,638 (26.6) & 2,273 (8.6)\\
         & Delayed, no. (\%) & 60 (0.6) & 0 (0.0)\\
        \multicolumn{2}{l}{Pancreatic tumor} & & &  \\
         & Yes, no. (\%) & 1,076 (10.8) & 2,829 (10.7) \\
         & No, no. (\%) & 8,825 (89.2) & 23,660 (89.3)\\
        \multicolumn{2}{l}{Dilated duct} & & & $8.0\times 10^{-1}$\\
         & Yes, no. (\%) & 4,164 (42.1) & 10,383 (39.2)\\
         & No, no. (\%) & 5,737 (57.9) & 16,106 (60.8)\\
        \multicolumn{2}{l}{Tumors per positive CT, no. (IQR)}             &  1.00 (1.00, 1.00) & 1.00 (1.00, 1.00) & $1.2\times 10^{-96}$\\
        \multicolumn{2}{l}{Tumor volume, mm$^{3}$ (IQR)}      &  2,927 (960, 8,084) & 2,996 (854, 15,158) & $2.0\times 10^{-4}$\\
        \multicolumn{2}{l}{Tumor HU value, mean (SD)}      & 60.8 (25.1) & 67.7 (56.4) & $4.1\times 10^{-2}$\\
        \multicolumn{2}{l}{Pancreas volume, mm$^{3}$ (IQR)}       &  80,653 (63,912, 98,315) & 74,850 (56,379, 92,378) & $3.9\times 10^{-79}$\\
        \multicolumn{2}{l}{Pancreas HU value, mean (SD)}      & 75.6 (35.8) & 85.6 (54.9) & $6.3\times 10^{-36}$\\
        \bottomrule
    \end{tabular}
\end{table}

\subsection{Dataset Contributors}

\begin{figure}[h]
\centering  
    \includegraphics[width=1\linewidth]{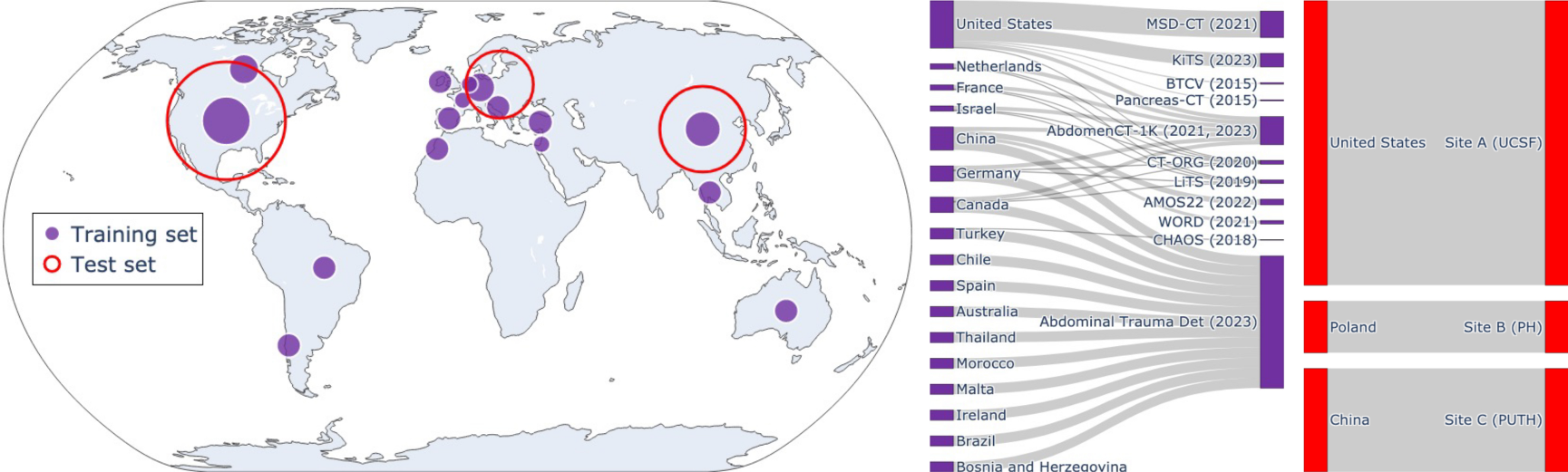}
    \caption{\textbf{Geographic diversity of the \dataset\ dataset.}
    Global distribution of contributing centers in the \dataset\ training set (purple circles) and test set (red outlines). Circle size is proportional to the base-10 logarithm (log$_{10}$) of the number of CT scans contributed per country. The training set is aggregated from diverse public datasets spanning multiple countries, while the much larger test set is exclusively drawn from three independent centers---UCSF (United States, North America), PH (Poland, Europe), and PUTH (China, Asia)---not seen during training, enabling rigorous out-of-distribution evaluation.
    }
    \label{fig:center_distribution}
\end{figure}

The CT scans for the \dataset\ dataset come from 145 centers in 18 countries. As summarized in Figure~\ref{fig:center_distribution}, the CT scans from the training set are assembled from 11 publicly available abdominal CT datasets; the test set are collected from 3 centers, including University of California, San Francisco (UCSF), Polish Hospitals (PH), and Peking University Third Hospital (PUTH).
All data are anonymized, and the CT scans have been reviewed visually to preclude the presence of personal identifiers. The only processing applied to the CT scans is a transformation into a unified NIfTI format using NiBabel in Python. All CT scans from the training set can be downloaded from their official websites; ethics approval was not required. The use of test set has received IRB approval from Johns Hopkins Medicine under IRB00403268.

\subsection{Annotation Protocol}

The pancreatic tumors in the \dataset\ dataset were manually annotated by a team of 23 medical annotators with varying levels of expertise in pancreatic imaging, as summarized in Table~\ref{tab:annotator_experience}. Each CT scan was annotated slice-by-slice using the MONAI-Label software \cite{cardoso2022monai}, with annotators assigning one of the pre-defined anatomical labels or marking the region as \textit{Background} if it did not correspond to any defined structure. Initial tumor annotations were performed by annotators with $\geq$3 years of radiology experience. Each annotation was then reviewed by three additional annotators who were blinded to the initial labels. In cases of disagreement, a specialist served as the final arbiter to resolve labeling conflicts. Extremely small or ambiguous lesion-like structures were excluded to ensure consistency and quality. This structured multi-annotator annotation process was designed to ensure consistency, resolve ambiguity, and achieve high-quality voxel-wise annotations. 

The \dataset\ dataset includes public organ and tumor segmentation datasets (Figure \ref{fig:center_distribution}). However, these datasets were not fully-annotated for all tumors and structures we have in \dataset. The public datasets inside the \dataset\ training set had 191 pancreatic tumor annotations. We annotated 885 additional pancreatic tumors, reaching 1,076 pancreatic tumor annotations in the \dataset\ training set. \appendixname~\ref{sec:supp_dataset_comparison} compares the number of structure annotations in public datasets and in \dataset. To efficiently scale voxel-wise annotations across pancreas head, body, tail, and 24 other anatomical structures, we employed a human-in-the-loop workflow \cite{qu2023annotating,li2025scalemai,zhang2024leveraging}. Specifically, an AI-based anatomy segmentator was used to generate initial organ annotations, which were then manually verified and corrected by radiologists. This AI-assisted workflow was used only for non-tumor structures; all pancreatic tumors were annotated and reviewed manually.

\begin{table}[t]
    \centering
    \caption{\textbf{Annotator experience.} The 23 medical annotators contributing to the \dataset\ dataset span a wide range of experience levels, with Specialists averaging 27 years of practice, General radiologists 10 years, and Residents 4 years. Despite this variation, the annotators interpret a high volume of CT scans annually—Specialists averaging $\sim$10,300/year, Generals $\sim$18,000/year, and Residents $\sim$16,000/year---ensuring both breadth and depth of radiological expertise across annotations. This mix of senior and junior readers supports consistent, high-quality labeling while enabling scalability across thousands of cases.
    }
    \label{tab:annotator_experience}
    \scriptsize
    \begin{tabular}{P{0.03\linewidth}P{0.14\linewidth}P{0.11\linewidth}P{0.1\linewidth}|P{0.03\linewidth}P{0.14\linewidth}P{0.11\linewidth}P{0.1\linewidth}}
        \toprule
        No. & Annotator ID & Experience (yr) & CT read / year & No. & Annotator ID & Experience (yr) & CT read / year \\
        \midrule
        1 & Specialist 1 (S1) & 24 & 12,000 & 2 & Specialist 2 (S2) & 22 & 12,000 \\
        3 & Specialist 3 (S3) & 35 & 8,000 & 4 & Specialist 4 (S4) & 30 & 8,000 \\
        5 & Specialist 5 (S5) & 28 & 9,000 & 6 & Specialist 6 (S6) & 19 & 13,000 \\
        7 & Specialist 7 (S7) & 23 & 11,000 & 8 & General 1 (G1) & 12 & 18,000 \\
        9 & General 2 (G2) & 8 & 18,000 & 10 & General 3 (G3) & 9 & 18,000 \\
        11 & General 4 (G4) & 10 & 18,000 & 12 & General 5 (G5) & 8 & 18,000 \\
        13 & General 6 (G6) & 13 & 18,000 & 14 & General 7 (G7) & 11 & 18,000 \\
        15 & General 8 (G8) & 10 & 18,000 & 16 & General 9 (G9) & 10 & 18,000 \\
        17 & General 10 (G10) & 13 & 18,000 & 18 & General 11 (G11) & 10 & 18,000 \\
        19 & Resident 1 (R1) & 5 & 16,000 & 20 & Resident 2 (R2) & 3 & 16,000 \\
        21 & Resident 3 (R3) & 4 & 16,000 & 22 & Resident 4 (R4) & 5 & 16,000 \\
        23 & Resident 5 (R5) & 5 & 16,000 \\
        \bottomrule
    \end{tabular}
\end{table}

\subsection{Annotation Standard}\label{sec:annotation_standard}

Tumor annotations include the entire pancreatic mass, incorporating both solid and cystic components as well as intralesional necrosis, while excluding adjacent organs, fat, and vasculature. The pancreatic parenchyma is annotated into head, body, and tail based on anatomical landmarks: the head includes the uncinate process, and extends up to the mesenteric vessels; the body-tail separation is set at about the midpoint between the mesenteric vessels and the end of the pancreas tail. Only glandular tissue is included, excluding surrounding fat, vessels, and the duodenum. The pancreatic duct is annotated as a low-attenuation tubular structure extending from the tail to the ampulla of Vater, including both the duct wall and lumen, but excluding adjacent parenchyma and vessels. Related abdominal vessels are annotated as follows: the celiac artery from its origin to its trifurcation; the superior mesenteric artery (SMA) from its aortic origin to the first major branch; the portal vein from the confluence with the splenic vein to its entry into the liver; and the splenic vein from the splenic hilum to its confluence with the portal vein. For all vessels, both lumen and wall are included, while surrounding fat, organs, and unrelated tissues are excluded. Annotation standards for other vessels, abdominal organs, thoracic structures, and skeletal landmarks are detailed in the \appendixname~\ref{sec:supp_annotation_standard}.

\begin{figure}[h]
\centering  
    \includegraphics[width=1.0\linewidth]{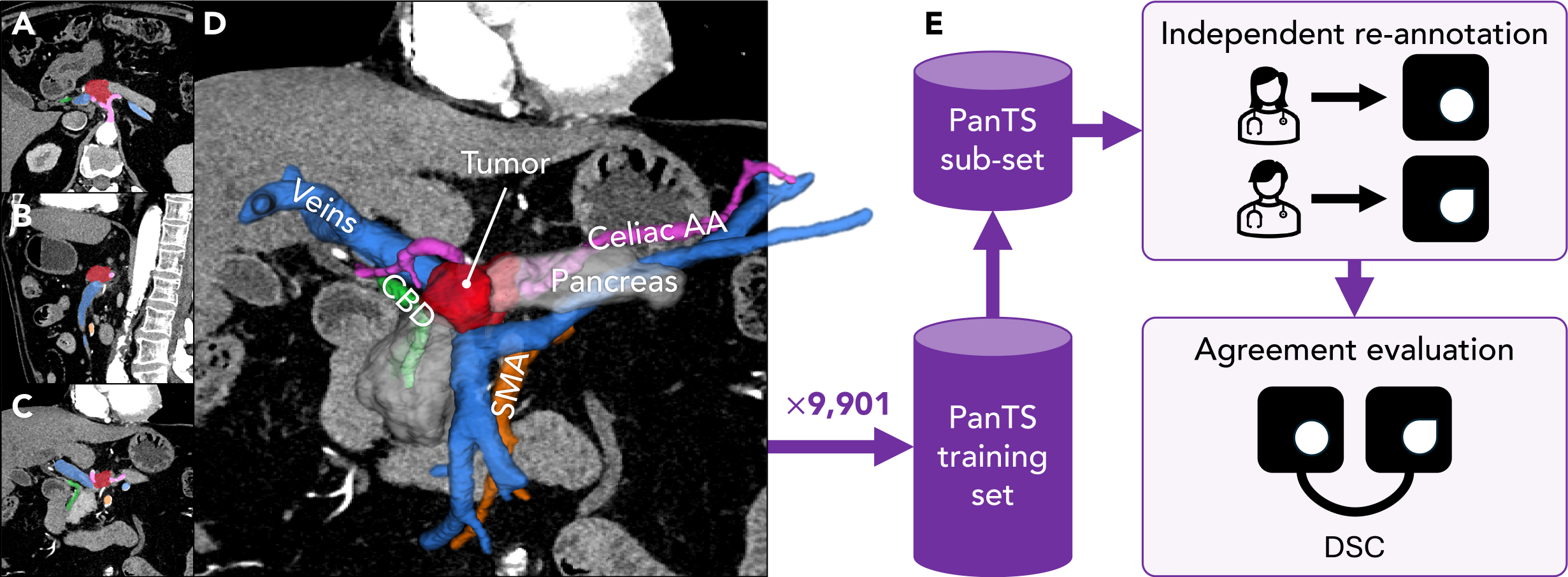}
    \caption{\textbf{Annotation standard and quality control.} \textbf{A--C.} Voxel-wise annotations of pancreatic tumors and surrounding anatomical structures shown on axial, sagittal, and coronal planes. Radiologists provide these annotations following the standard described in \S\ref{sec:annotation_standard}. \textbf{D.} 3D rendering on the coronal plane highlights detailed annotations of the tumor, pancreas, and key vessels, including the celiac artery (Celiac AA), superior mesenteric artery (SMA), common bile duct (CBD), and surrounding veins. \textbf{E.} To assess annotation quality, a subset of 300 CT scans from the \dataset\ training set was independently re-annotated by multiple radiologists. Inter-annotator agreement was evaluated using the Dice Similarity Coefficient (DSC).
    }
    \label{fig:annotation_standard_quality_control}
\end{figure}

\begin{figure}[t]
\centering  
    \includegraphics[width=1.0\linewidth]{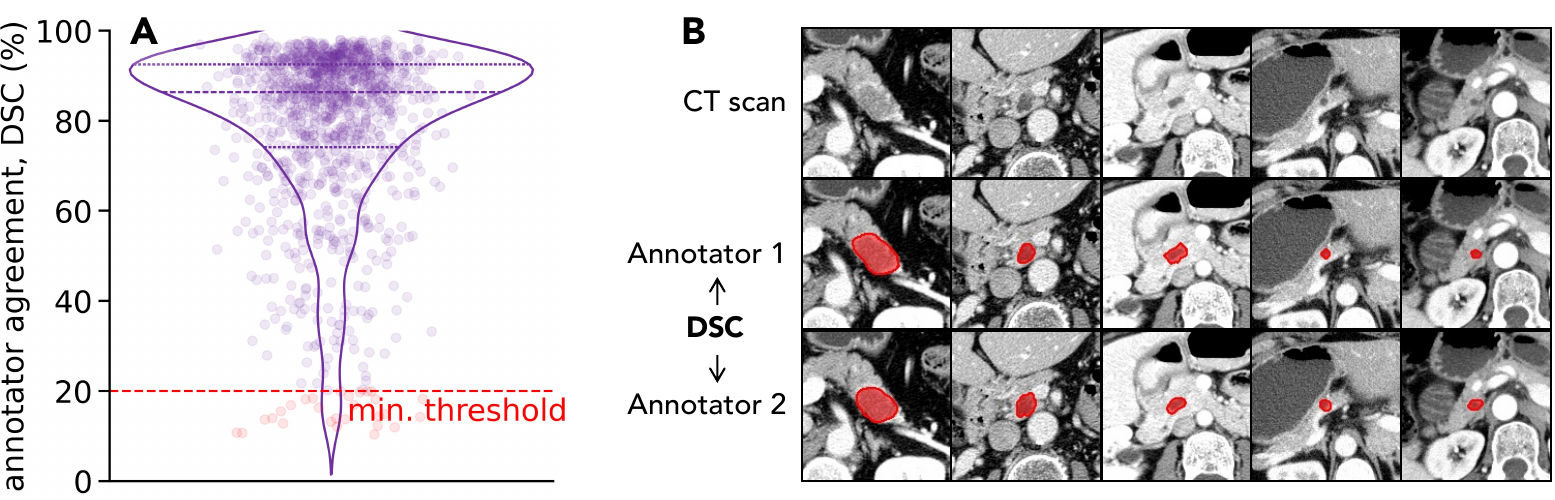}
    \caption{\textbf{Inter-annotator agreement on the \dataset\ subset.} \textbf{A.} Distribution of DSC (\%) values between two independent radiologists across 300 CT scans from the \dataset\ training set. Most annotations demonstrate high agreement, confirming their reliability. A minimum threshold of DSC = 20\% (dashed red line) is used to flag low-agreement cases, which are reviewed by senior radiologists for further quality assurance. \textbf{B.} Representative examples showing the same CT scan annotated by two different radiologists. High-agreement cases appear in the left columns, while low-agreement cases—often involving small or ambiguous lesions—appear on the right. 
    }
    \label{fig:interannotator_agreement}
\end{figure}

\subsection{Annotation Quality Control}

Large medical image datasets inevitably contain annotation imperfections, particularly in voxel-wise annotations. While such datasets remain highly valuable, their utility can be further enhanced by systematically assessing annotation reliability. To evaluate internal consistency and quality of voxel-wise annotations in our training set, we conducted an inter-annotator agreement study (Figure~\ref{fig:annotation_standard_quality_control}E).

Specifically, we randomly selected 300 CT scans from the training set and had them independently re-annotated by a second radiologist, blind to the initial annotation. We computed the Dice Similarity Coefficient (DSC) between the two annotations for each case as a measure of agreement (Figure~\ref{fig:interannotator_agreement}A). The median inter-annotator agreement was DSC (\%) = 86.1\%, with an interquartile range (IQR) of 19.6\%, indicating high consistency across annotators. However, a small number of cases showed low agreement (DSC $<$ 20\%), often due to small or ambiguous lesions. To ensure the annotation quality, we define a minimum threshold of DSC = 20\% and flag all such cases for review and possible correction by senior radiologists.

Figure~\ref{fig:interannotator_agreement}B shows representative examples of CT scans annotated by two radiologists. High-agreement cases are shown on the left, while low-agreement cases---typically more subtle or ambiguous---are shown on the right. This inter-annotator evaluation not only ensures annotation quality control but also provides a reference for benchmarking automated models: systems that achieve DSCs comparable to or exceeding this agreement level can be considered human-comparable in segmentation performance.

\section{Justification of Annotating Large-Scale Tumor Datasets}\label{sec:scaling}

A central hypothesis is that scaling up voxel-wise tumor annotations significantly improves AI performance, particularly under out-of-distribution (OOD) settings---like hospitals not seen in training. To evaluate this, we trained a standard nnU-Net model on pancreatic tumor datasets of increasing size---MSD-Pancreas ($n$ = 281), PANORAMA ($n$ = 2,238), and our proposed \dataset\ dataset ($n$ = 9,901)---and evaluated detection performance on the held-out \dataset\ test set, which contains CT scans from medical centers not present in any training data.

\begin{figure}[h]
\centering  
    \includegraphics[width=1\linewidth]{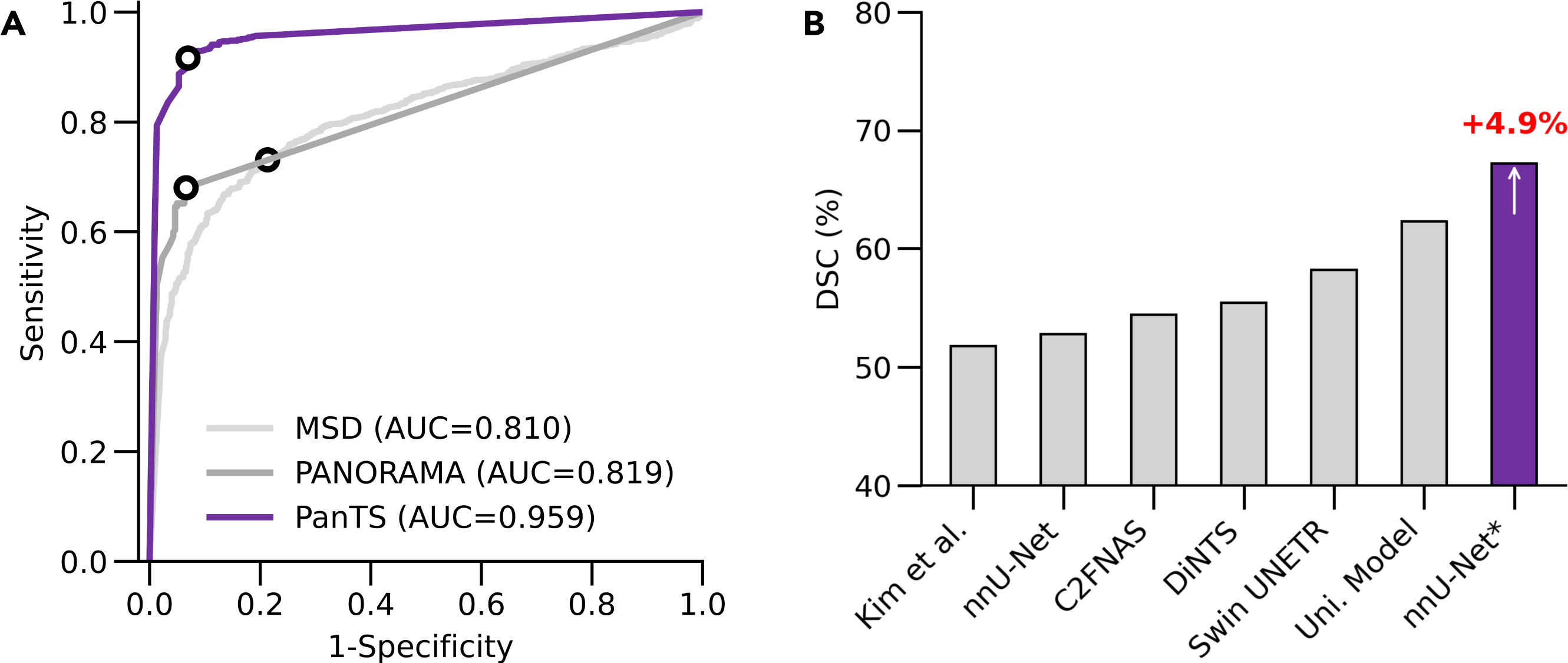}
    \caption{\textbf{Justification of annotating large-scale tumor datasets.}
    \textbf{A.} The Receiver Operating Characteristic (ROC) curve of standard nnU-Net trained on different scale of pancreatic CT datasets, i.e., MSD-Pancreas ($n$ = 281), PANORAMA ($n$ = 2,238), and our \dataset\ dataset ($n$ = 9,901). The performance is tested on the \dataset\ test dataset (CT collected different centers from MSD-Pancreas, PANORAMA, and the \dataset\ training set, detailed in Figure~\ref{fig:center_distribution}). The observation is the larger training set, the better pancreatic tumor detection performance on the out-of-distribution test set. \textbf{B.} Barplot of AI trained on our \dataset\ vs. AI trained on publicly available dataset (MSD-Pancreas). The performance is tested on the official MSD-Pancreas test set (third-party evaluation). All metrics can be found at \href{https://decathlon-10.grand-challenge.org/evaluation/challenge/leaderboard/}{The MSD Leaderboard}.
    } \label{fig:justification_large_scale}
\end{figure}

As shown in Figure~\ref{fig:justification_large_scale}A, model performance improves with dataset scale, but not uniformly. The Area Under the ROC Curve (AUC) increases modestly from 0.810 (MSD) to 0.819 (PANORAMA), and then substantially to 0.959 when trained on our \dataset\ dataset\footnote{We hypothesize this discrepancy stems from annotation protocol differences: PANORAMA only annotates pancreatic ductal adenocarcinoma (PDAC), while treating all other tumors and healthy pancreases as \textit{Normal}. This conflates distinct conditions under a single label, introducing ambiguity and limiting the model’s ability to learn fine-grained distinctions between normal and abnormal tissue.}. While this trend partially aligns with AI scaling laws \cite{kaplan2020scaling,zhai2022scaling}---which suggest that performance improves logarithmically with dataset size---the limited gain from MSD to PANORAMA indicates that scale alone is not sufficient. The significant improvement observed with \dataset\ is instead attributable to both its larger size and its high-quality, comprehensive annotations. \dataset\ includes 9,901 CT scans from 145 centers, capturing a broad range of pancreatic tumor types, anatomical variations, scan protocols, and noise distributions---factors essential for building robust, generalizable AI models.

To further assess the benefit of large-scale annotation, we benchmark nnU-Net trained on our \dataset\ dataset against leading AI methods trained on MSD (Figure~\ref{fig:justification_large_scale}B). Using the official MSD test set, and third-party evaluated by the organizers of MSD challenge, our nnU-Net trained on \dataset\ outperforms all baseline methods by a margin of at least +4.9\% DSC and +3.1\% NSD in pancreatic tumor segmentation, becoming the new top-1 AI model in the public MSD-Pancreas leaderboard.

\section{Justification of Annotating 24 Surrounding Anatomical Structures}\label{sec:24_anotamical_structure}

\begin{figure}[h]
\centering  
    \includegraphics[width=1\linewidth]{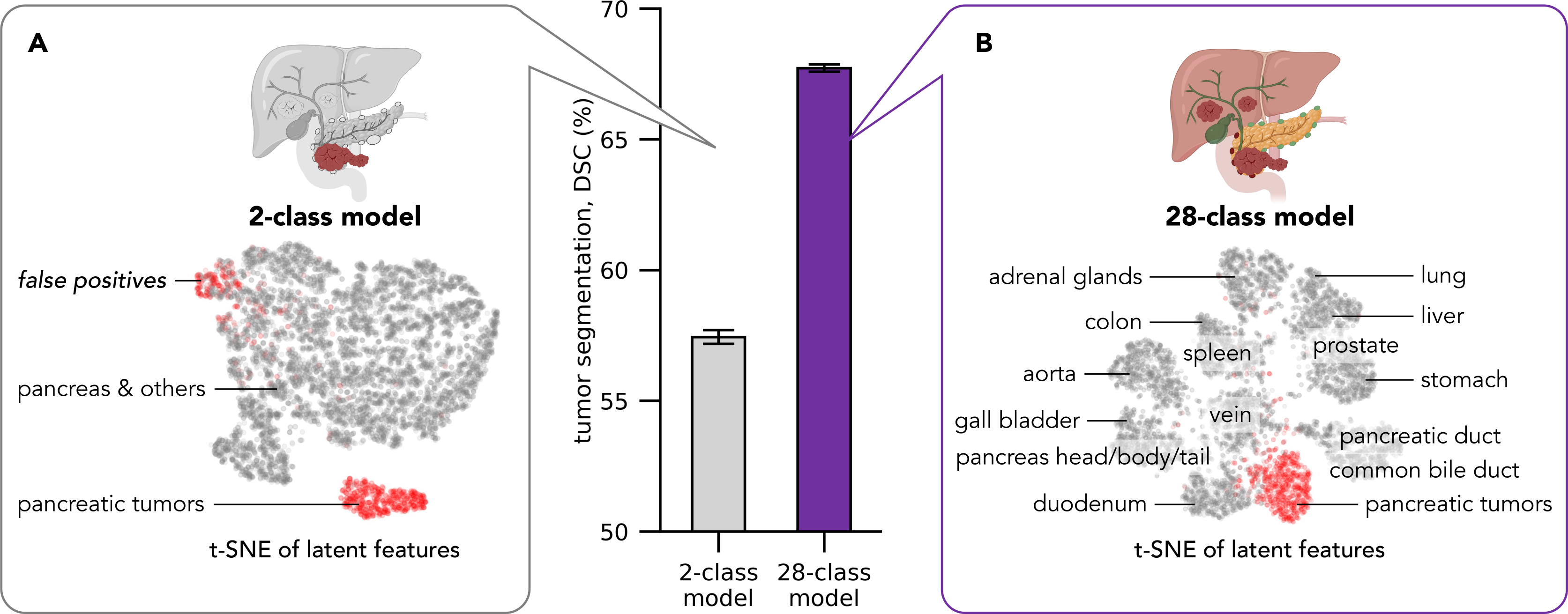}
    \caption{\textbf{Justification of annotating 24 surrounding anatomical structures.} 
    We compare nnU-Net models trained with 2 classes (tumor and pancreas) versus 28 classes (tumor, pancreas head/body/tail, and 24 surrounding anatomical structures). The 28-class model significantly improves tumor segmentation accuracy (mean DSC +10.3\%, p < 0.0001), highlighting the value of anatomical context. We further analyze the latent features of the two nnU-Net models. \textbf{A.} The 2-class model, trained to distinguish only pancreatic tumors vs. background, shows overlapping feature clusters in t-SNE space \cite{van2008visualizing}, with substantial false positives. \textbf{B.} The 28-class model, trained with supervision from 27 additional anatomical structures, results in better separation of pancreatic tumor features from surrounding tissues in t-SNE space.
    }
    \label{fig:justification_more_structures}
\end{figure}

To assess the impact of anatomical context on pancreatic tumor segmentation, we compared the performance of a standard nnU-Net trained under two labeling schemes: a 2-class setup (tumor and pancreas) and a 28-class setup (tumor, pancreas subregions—head, body, tail—and 24 surrounding anatomical structures). Figure~\ref{fig:justification_more_structures}A shows the 28-class model markedly outperforms the 2-class model in tumor segmentation, with mean DSC improving \textbf{+10.3\%} from 57.4\% to 67.7\%. Tumor boundary accuracy, measured by Normalized Surface Dice (NSD), also increases \textbf{+9.7\%} from 56.8\% to 66.5\%.

By including structures such as the duodenum, bile duct, and nearby vessels, the 28-class model leverages additional spatial context to more effectively exclude non-tumorous tissue near ambiguous boundaries, enhancing spatial reasoning in anatomically complex regions. Annotating adjacent organs further encourages the model to internalize critical spatial relationships, especially in areas with low-contrast boundaries \cite{kang2023label,zhu2022covid}. These findings suggest that anatomical annotations function as implicit regularizers, helping the model structure its latent space more effectively.

The addition of 24 surrounding structures provides vital contextual cues, enabling clearer differentiation of tumors from neighboring tissues. This enriched anatomical supervision guides the model to learn spatial relationships, structural boundaries, and typical organ configurations—particularly important in the pancreas. These results highlight the importance of comprehensive multi-organ annotation for training robust and generalizable AI models in medical imaging.

In summary, our results confirm that including spatially related anatomical structures can improve segmentation of the class of interest. This underscores the importance of extensive anatomical annotation when designing large-scale, high-performance medical AI datasets.

\section{Conclusion and Discussion}
\label{sec:conclusion}

Our \dataset\ dataset represents a major leap forward in data-driven pancreatic cancer research. With over 36,000 CT scans from 145 medical centers and nearly one million expert-validated voxel-wise annotations, \dataset\ is the largest and most diverse publicly available resource for pancreatic tumor analysis to date. This dataset was developed through an immense collaborative effort involving 23 radiologists and years of manual annotation, quality assurance, and cross-validation.

Despite its strengths, \dataset\ highlights the considerable challenges of annotating tumor datasets compared to normal anatomical structures. Even among experts, inter-annotator agreement can be modest, especially for small, ambiguous lesions. Our analysis of misclassified cases provides insight: in false positives, annotators noted subtle texture irregularities in the pancreas but without the hallmark signs of tumor presence (e.g., ductal dilation or parenchymal atrophy). Conversely, false negatives often involved subtle or atypical presentations, such as exophytic growths in hard-to-visualize regions as the pancreas tail or diffuse parenchymal thinning that may indicate underlying malignancy.

These findings underscore a central challenge: even experienced radiologists can miss early or atypical tumors, emphasizing the potential value of AI models trained on large, richly annotated datasets like \dataset. At the same time, they highlight the need for caution when interpreting both manual and automated annotations---especially in edge cases. Future work should explore multimodal learning, combining imaging, pathology, and clinical data, to further improve accuracy and reduce uncertainty.

Importantly, \dataset\ is more than a technical benchmark---it has clinical and translational significance. Pancreatic cancer remains one of the deadliest malignancies due to late-stage diagnoses and the subtlety of early radiologic signs. While AI holds promise for earlier detection, prior models have been hampered by small, homogeneous training data. By contrast, \dataset\ offers unprecedented scale and diversity, enabling the development of robust, generalizable AI systems. It also provides a foundation for anatomy-aware evaluation metrics, automated report generation, subpopulation analysis, and AI-assisted education. To maximize impact, we publicly release the baseline model and the \dataset\ training set under a non-commercial license, and we release benchmarking protocols to allow easy third-party evaluation in the \dataset\ test set.

\begin{ack}
This work was supported by the Lustgarten Foundation for Pancreatic Cancer Research, the Patrick J. McGovern Foundation Award, and the National Institutes of Health (NIH) under Award Number R01EB037669. We would like to thank the Johns Hopkins Research IT team in \href{https://researchit.jhu.edu/}{IT@JH} for their support and infrastructure resources where some of these analyses were conducted; especially \href{https://researchit.jhu.edu/research-hpc/}{DISCOVERY HPC}.
\end{ack}

\bibliographystyle{plainnat}
\bibliography{zzhou,refs}

\clearpage

\appendix

\renewcommand \thepart{}
\renewcommand \partname{}

\part{Appendix} 
\setcounter{secnumdepth}{4}
\setcounter{tocdepth}{4}
\parttoc 

\clearpage
\section{Related Datasets \& Our Contribution}\label{sec:supp_dataset_comparison}

{\setlength{\tabcolsep}{1.4pt}%
 \renewcommand{\arraystretch}{0.96}%
 \begin{table}[h]
\caption{\textbf{Comparison of \dataset\ with public abdominal CT datasets.} This comparative summary underscores the breadth, depth, and clinical relevance of \dataset\ relative to existing public datasets. While a number of prior datasets were incorporated into our training partition, our team made substantial and transformative contributions. Specifically, 23 board-certified radiologists independently annotated and rigorously validated previously unlabeled pancreatic tumors as well as over 25 additional abdominal and thoracic anatomical structures, many of which were not comprehensively labeled in the source datasets. This effort significantly elevates the clinical utility and completeness of the dataset. 
\textbf{Scale:} With 36,390 CT scans, \dataset\ is over 8.5$\times$ larger than the most extensive existing dataset dedicated to pancreatic tumor detection, setting a new benchmark for scale in abdominal imaging datasets.  
\textbf{Quality:} All tumor annotations meet silver-standard criteria, with expert oversight ensuring high inter-rater reliability and consistency.  
\textbf{Diversity:} The scans were collected from 145 institutions spanning 20 countries, offering a level of demographic and scanner variability that is 3$\times$ more diverse than previous benchmarks—critical for training generalizable and robust AI models.  
Collectively, these attributes make \dataset\ one of the most comprehensive, diverse, and clinically curated resources available for abdominal imaging research. \textit{To advance transparency, reproducibility, and real-world relevance, we will publicly release the \dataset\ training set and use the \dataset\ test set to benchmark the performance of AI algorithms.}
 }
 \scriptsize
 \begin{tabular*}{\textwidth}{@{\extracolsep{\fill}}
   p{0.175\linewidth}  
   W                   
   W                   
   P{0.025\linewidth}   
   P{0.025\linewidth}   
   *{25}{C}}           
 \toprule
 dataset &
 \rotatebox{90}{pancreatic\,tumors} &
 \rotatebox{90}{number of CTs} &
 \rotatebox{90}{institutions} &
 \rotatebox{90}{countries} &
 \rotatebox{90}{pancreas} &
 \rotatebox{90}{pancr. head/body/tail} &
 \rotatebox{90}{SMA} &
 \rotatebox{90}{pancreatic\,duct} &
 \rotatebox{90}{celiac\,artery} &
 \rotatebox{90}{CBD} &
 \rotatebox{90}{pancreatic veins} &
 \rotatebox{90}{aorta} &
 \rotatebox{90}{gallbladder} &
 \rotatebox{90}{L\,kidney} &
 \rotatebox{90}{R\,kidney} &
 \rotatebox{90}{liver} &
 \rotatebox{90}{IVC} &
 \rotatebox{90}{spleen} &
 \rotatebox{90}{stomach} &
 \rotatebox{90}{L\,adrenal} &
 \rotatebox{90}{R\,adrenal} &
 \rotatebox{90}{bladder} &
 \rotatebox{90}{colon} &
 \rotatebox{90}{duodenum} &
 \rotatebox{90}{L\,femur} &
 \rotatebox{90}{R\,femur} &
 \rotatebox{90}{L\,lung} &
 \rotatebox{90}{R\,lung} &
 \rotatebox{90}{prostate}\\
\midrule
KiTS'23 \citeyearpar{heller2020international}                & 0   & 489   & 1  & 1  & 
\no&\no&\no&\no&\no&\no&\no&\no&\no&\yes&\yes&\no&\no&\no&\no&\no&\no&\no&\no&\no&\no&\no&\no&\no&\no\\
LiTS \citeyearpar{bilic2019liver}                            & 0   & 131   & 7  & 5  & 
\no&\no&\no&\no&\no&\no&\no&\no&\no&\no&\no&\yes&\no&\no&\no&\no&\no&\no&\no&\no&\no&\no&\no&\no&\no\\
TCIA-Pancr.-CT \citeyearpar{roth2015deeporgan}             & 0   & 42    & 1  & 1  & 
\yes&\no&\no&\no&\no&\no&\no&\no&\no&\no&\no&\no&\no&\no&\no&\no&\no&\no&\no&\no&\no&\no&\no&\no&\no\\
CT-ORG \citeyearpar{rister2020ct}                            & 0   & 140   & 8  & 6  & 
\no&\no&\no&\no&\no&\no&\no&\no&\no&\yes&\yes&\yes&\no&\no&\no&\no&\no&\yes&\no&\no&\yes&\yes&\yes&\yes&\no\\
Trauma Det. \citeyearpar{rsna-2023-abdominal-trauma-detection} & 0   &4,714 &23  &13  & 
\no&\no&\no&\no&\no&\no&\no&\no&\no&\yes&\yes&\yes&\no&\yes&\yes&\no&\no&\no&\yes&\yes&\no&\no&\no&\no&\no\\
BTCV \citeyearpar{landman2015miccai}                         & 0   & 47    & 1  & 1  & 
\yes&\no&\no&\no&\no&\no&\yes&\yes&\yes&\yes&\yes&\yes&\yes&\yes&\yes&\yes&\yes&\no&\no&\no&\no&\no&\no&\no&\no\\
CHAOS \citeyearpar{valindria2018multi}                       & 0   & 20    & 1  & 1  & 
\no&\no&\no&\no&\no&\no&\no&\no&\no&\no&\no&\yes&\no&\no&\no&\no&\no&\no&\no&\no&\no&\no&\no&\no&\no\\
AbdomenCT-1K \citeyearpar{ma2021abdomenct}                   & 0   &1,050 &12  & 7  & 
\yes&\no&\no&\no&\no&\no&\no&\no&\no&\yes&\yes&\yes&\no&\yes&\no&\no&\no&\no&\no&\no&\no&\no&\no&\no&\no\\
WORD \citeyearpar{luo2021word}                               & 0   & 120   & 1  & 1  & 
\yes&\no&\no&\no&\no&\no&\no&\no&\yes&\yes&\yes&\yes&\no&\yes&\yes&\yes&\yes&\yes&\yes&\yes&\yes&\yes&\no&\no&\no\\
AMOS \citeyearpar{ji2022amos}                                & 0   & 200   & 2  & 1  & 
\yes&\no&\no&\no&\no&\no&\no&\yes&\yes&\yes&\yes&\yes&\yes&\yes&\yes&\yes&\yes&\yes&\no&\yes&\no&\no&\no&\no&\yes\\
MSD-CT \citeyearpar{antonelli2021medical}                    &191  & 945   & 1  & 1  & 
\yes&\no&\no&\no&\no&\no&\no&\no&\no&\no&\no&\yes&\no&\yes&\no&\no&\no&\no&\yes&\no&\no&\no&\yes&\yes&\no\\
PANORAMA \citeyearpar{alves2024panorama} & 578 & 2,238 & 7 & 1 & \yes&\no&\no&\no&\no&\no&\no&\no&\no&\no&\no&\no&\no&\no&\no&\no&\no&\no&\no&\no&\no&\no&\no&\no&\no\\
\midrule
\textbf{\dataset\ (ours)} \\
training set & 1,076   &  9,901  & 145   &  17  & \yes &\yes
   & \yes  & \yes  &  \yes & \yes  & \yes  & \yes  & \yes  & \yes  & \yes  &  \yes &  \yes &  \yes & \yes  &\yes   & \yes  &\yes   &\yes   &\yes   &\yes   &\yes   &\yes   &\yes   &\yes\\
test set & 2,829   &  26,489  & 3   &  3  & \yes &\yes
   & \yes  & \yes  &  \yes & \yes  & \yes  & \yes  & \yes  & \yes  & \yes  &  \yes &  \yes &  \yes & \yes  &\yes   & \yes  &\yes   &\yes   &\yes   &\yes   &\yes   &\yes   &\yes   &\yes\\
\bottomrule
\end{tabular*}
\end{table}
}

\clearpage
\section{Baseline and Implementation Details}\label{sec:supp_implementation_details}

\subsection{Top-Performing Methods in Medical Segmentation Decathlon}

\textbf{Kim~\etal}~\cite{kim2019scalable} proposed a neural architecture search (NAS) framework for 3D medical image segmentation tasks. This method explores a broad design space by automatically searching for optimal layer-wise structures, including both neural connectivities and operation types, across the encoder and decoder stages. To address the high computational cost associated with high-resolution 3D data, the framework introduces a scalable stochastic sampling algorithm based on continuous relaxation, which enables efficient gradient-based optimization.
 
\textbf{nnU-Net}~\cite{isensee2021nnu,isensee2024nnu} is a self-configuring segmentation framework. It automatically configures pre-processing, network architecture, training, and post-processing. Its auto-configuration is guided by a combination of fixed parameters, interdependent rules that account for dataset characteristics and computational constraints, as well as empirical heuristics.

\textbf{C2FNAS}~\cite{yu2020c2fnas}  is a coarse-to-fine neural architecture search (C2FNAS) framework designed to reduce the complexity and manual effort involved in developing task-specific 3D segmentation networks. This method addresses the common issue of inconsistency between the search and deployment stages in traditional NAS—often caused by memory limitations and expansive search spaces—by decoupling the architecture search into two successive phases. In the coarse stage, the framework explores the macro-level network topology, determining how convolutional modules are connected. In the fine stage, it refines the architecture by selecting specific operations within each cell, guided by the previously discovered topology. This coarse-to-fine strategy mitigates search-deployment mismatches while preserving scalability.

\textbf{DiNTS}~\cite{he2021dints} introduces a differentiable neural architecture search (NAS) framework tailored for 3D medical image segmentation, which aims to enable flexible topology design, high search efficiency, and controlled GPU memory usage. Unlike traditional NAS methods that are constrained by fixed topologies (e.g., U-Net) or suffer from long search times on large 3D datasets, DiNTS facilitates the automatic discovery of multi-path network topologies through a highly flexible and continuous search space. To address the discretization gap—the performance drop observed when converting an optimal continuous architecture into a discrete one—the method incorporates a topology loss to preserve the quality of the searched architecture. Furthermore, DiNTS integrates GPU memory constraints directly into the search process, making it more practical for resource-intensive 3D tasks.

\textbf{Swin UNETR}~\cite{tang2022self} adapted Swin Transformers to enhance medical image segmentation by capturing both local and global features through a hierarchical, window-based self-attention mechanism, outperforming the original UNETR by effectively modeling global context with Swin Transformers. Additionally, self-supervised pre-training of Swin Transformers on large-scale unlabeled 3D medical image datasets—using techniques such as masked autoencoding—can significantly boost model robustness and downstream task performance. These features led to state-of-the-art performance in various 3D medical image analysis applications, particularly in CT segmentation tasks.

\textbf{Universal Model}~\cite{liu2023clip,liu2024universal,zhang2023continual} was proposed to overcome the limitations of dataset-specific models in organ and tumor segmentation. Traditional models often suffer from poor generalizability due to the small size, partial annotations, and limited diversity of individual datasets. In contrast, the proposed model leverages text embeddings derived from Contrastive Language-Image Pre-training (CLIP) to encode anatomical labels. This enables the model to learn semantically structured feature representations and facilitates the segmentation of 25 organs and 6 tumor types across diverse anatomical regions. The model demonstrates strong transferability to novel domains and previously unseen tasks.

\subsection{Experimental Setting}

\subsubsection{Justification of Annotating Large-Scale Tumor Datasets} To verify the effectiveness of scaling up voxel-wise tumor annotations and to justify the annotation of the \dataset\ dataset, we designed two comparative experiments to assess how increasing the volume of annotated data affects model performance, particularly in out-of-distribution (OOD) scenarios. 

\begin{itemize}
    \item Experiment 1: We selected two widely used public datasets—MSD-Pancreas ($n$ = 281) and PANORAMA ($n$ = 2,238)—as representative baselines for comparison with our proposed large-scale dataset, \dataset\ ($n$ = 9,901). A standard nnU-Net model was independently trained on each of the three datasets using identical configurations, including network architecture, data preprocessing, augmentation strategies, and optimization parameters, to ensure a fair comparison. All models were evaluated on the \dataset\ test set, which consists of CT scans from medical centers not included in the training data.

    \item Experiment 2: We benchmarked nnU-Net trained on the \dataset\ dataset against leading AI methods trained on the MSD dataset. Specifically, we selected Kim~\etal, nnU-Net, C2FNAS, DiNTS, Swin UNETR, and Uni. Model as baselines for comparison, all trained on the MSD training set. The official MSD test set was used for evaluation, with performance independently evaluated by the organizers of the MSD challenge.
\end{itemize}

This experimental setting enables quantification of the benefits of large-scale tumor annotation by comparing model performance across datasets of increasing size and by evaluating under both in-distribution and out-of-distribution conditions.

\subsubsection{Justification of Annotating 24 Surrounding Anatomical Structures}

To evaluate whether incorporating detailed anatomical context improves the ability of tumor segmentation models to distinguish tumor boundaries, we conducted a comparative study under two labeling schemes. The core hypothesis is that segmenting additional surrounding structures enables the network to better capture anatomical boundaries and spatial relationships, thereby enhancing its ability to localize and delineate tumors.

Specifically, we trained the standard nnU-Net model using two distinct annotation protocols:
\begin{itemize}
    \item A 2-class setup, including only the tumor and pancreas regions, reflecting the minimal annotation approach commonly used in public datasets.

    \item A 28-class setup, encompassing the tumor, pancreas subregions (head, body, and tail), and 24 surrounding anatomical structures, including vessels, gastrointestinal organs, and adjacent tissues.
\end{itemize}

Both models were trained on the same cohort of CT scans from the \dataset\ dataset, ensuring that performance differences are solely attributable to the inclusion of more comprehensive structural annotations. All training configurations—including preprocessing steps, augmentation strategies, and optimization parameters—were held constant across both setups. By comparing segmentation results on the held-out \dataset\ test set, we assessed whether finer-grained anatomical annotations enhance generalization performance and tumor localization accuracy.

\subsection{Implementation Details}

\subsubsection{Justification of Annotating Large-Scale Tumor Datasets.}

\begin{itemize}
    \item Experiment 1: The three standard nnU-Net models were trained using the nnU-Net framework. The orientation of CT scans was standardized to a consistent anatomical orientation. All preprocessing parameters—including resampling spacing, intensity range, and crop size—were automatically selected by the nnU-Net framework through empirical optimization on each training dataset. Detailed configuration settings are included in the accompanying code repository as JSON files. Data augmentation during training followed the default strategies defined by the nnU-Net framework. All models were trained for 1,000 epochs, each consisting of 250 iterations. We employed the SGD optimizer with a base learning rate of 0.01 and a batch size of 2. During inference, we applied test-time augmentation and used the sliding window strategy with an overlap ratio of 0.5, following the default nnU-Net implementations.

    \item Experiment 2: The training and inference procedures for our nnU-Net model followed the same configurations described in Experiment 1. For the comparative models, we report the official results released by the MSD Challenge organizers on the public leaderboard.
\end{itemize}

\subsubsection{Justification of Annotating Large-Scale Tumor Datasets.}
The two standard nnU-Net models were trained using the nnU-Net framework, following training procedures consistent with those described in Experiment 1. The only distinction between the two setups lies in the class labels used for training, with all other configurations kept identical.

\subsection{Evaluation Metrics}
Each evaluation metric captures a specific aspect of the results, and selecting appropriate metrics is essential to highlight the characteristics of interest. To quantitatively evaluate segmentation performance, we employ a suite of widely adopted metrics: Dice Similarity Coefficient (DSC), Normalized Surface Dice (NSD), Sensitivity, Specificity, and Area Under the Receiver Operating Characteristic Curve (AUC).

\subsubsection{Dice Similarity Coefficient (DSC)} 

DSC measures the volumetric overlap between the predicted segmentation and the ground truth. It is defined as:
\begin{equation}
\text{DSC} = \frac{2 |P \cap G|}{|P| + |G|}
\end{equation}
where $P$ and $G$ denote the sets of predicted and ground truth positive voxels, respectively. DSC ranges from 0 to 1, with higher values indicating better agreement. It is particularly useful for handling imbalanced data and is the standard metric in many medical imaging tasks.

\subsubsection{Normalized Surface Dice (NSD)} 

NSD evaluates the agreement between the predicted and ground truth surfaces within a specified tolerance $\tau$, which reflects clinically acceptable deviation. It is defined as:
\begin{equation}
\text{NSD} = \frac{|\{x \in \partial P : \exists y \in \partial G, \|x - y\| < \tau\}| + |\{y \in \partial G : \exists x \in \partial P, \|y - x\| < \tau\}|}{|\partial P| + |\partial G|},
\end{equation}
where $\partial P$ and $\partial G$ represent the surfaces of the predicted and ground truth segmentations. NSD provides a more stringent surface-level evaluation, which is especially relevant in clinical applications requiring precise boundary delineation.

\subsubsection{Sensitivity \& Specificity} 

Sensitivity (also known as recall or true positive rate) quantifies the proportion of actual positives correctly identified, while Specificity measures the proportion of actual negatives correctly identified. They are defined as:
\begin{equation}
\text{Sensitivity} = \frac{TP}{TP + FN}, \quad \text{Specificity} = \frac{TN}{TN + FP},
\end{equation}
where $TP$, $TN$, $FP$, and $FN$ are the numbers of true positives, true negatives, false positives, and false negatives, respectively. High sensitivity is critical for minimizing missed detections, whereas high specificity is important to reduce false alarms.

\subsubsection{Area Under the Receiver Operating Characteristic Curve (AUC)} 

The AUC quantifies the overall ability of a model to discriminate between classes by measuring the area under the ROC curve, which illustrates the trade-off between sensitivity and specificity across varying thresholds. The ROC curve plots sensitivity against (1 – specificity) across different threshold values. An AUC value of 1.0 indicates perfect classification, while a value of 0.5 represents random guessing. AUC is particularly useful for evaluating the model’s discriminative capability in segmentation tasks.

\clearpage
\section{Annotation Standard}\label{sec:supp_annotation_standard}

\textbf{Pancreas and Related Structures.} \textit{Pancreatic tumors:} Annotate the entire tumor mass regardless of location within the pancreas. Include both solid and cystic components, as well as any intralesional necrosis. Exclude adjacent organs, fat, and vasculature. \textit{Pancreas head, body, and tail:} Annotate the pancreatic parenchyma divided into three anatomical regions. The head is located to the right of the superior mesenteric vessels, within the curvature of the duodenum, and includes the uncinate process. The body lies between the left border of the superior mesenteric vessels and the left edge of the aorta. The tail lies anterior to the aorta, extending toward the splenic hilum. Include the entire gland parenchyma, excluding surrounding fat, vessels, and the duodenum. \textit{Pancreatic duct:} Identify as a low-attenuation tubular structure within the pancreas. Annotate from the tail to the ampulla of Vater, including both the duct wall and lumen. Exclude surrounding pancreatic parenchyma and vessels.

\textbf{Vascular Structures.} \textit{Aorta:} Annotate the entire lumen from the diaphragm to the bifurcation. Include the arterial wall and any calcifications, ulcers, thrombus, or dissection. Exclude surrounding tissues and organs. \textit{Celiac artery:} Identify as a short arterial branch from the aorta. Annotate from its origin to its division into the left gastric, splenic, and common hepatic arteries. Include the lumen and wall. Exclude surrounding fat and organs. \textit{Superior mesenteric artery (SMA):} Trace from its origin at the aorta to the point of major branching. Include the vessel wall and lumen. Exclude surrounding fat, pancreas, and bowel. \textit{Postcava:} Annotate the entire lumen and wall from its origin at the postcava to its entry into the right atrium. Include any intraluminal thrombus. Exclude surrounding fat and structures. \textit{Portal vein:} A bright, enhanced vessel formed by the confluence of the SMV and splenic vein. Annotate from the confluence to liver entry. Include lumen, wall, and any thrombus. \textit{Splenic vein:} Trace from the spleen to its confluence with the SMV. Include lumen and wall, excluding adjacent pancreatic tissue and fat.

\textbf{Abdominal Organs.} \textit{Liver:} Annotate the entire parenchyma including all segments, intrahepatic vessels, bile ducts, and any hepatic lesions. Exclude adjacent organs and fat. \textit{Spleen:} Annotate the entire splenic parenchyma and any lesions. Exclude surrounding fat and nearby structures such as stomach, kidney, and colon. \textit{Left and right kidneys:} Annotate the renal parenchyma. Exclude renal pelvis, ureter, perirenal fat, and adjacent structures. Include renal lesions if present. \textit{Left and right adrenal glands:} Annotate the entire gland and any lesions. Exclude surrounding fat and nearby organs. \textit{Gall bladder:} Annotate the wall and lumen, including the fundus, body, and neck. Include gallstones or polyps. Exclude cystic duct and liver tissue. \textit{Stomach:} Annotate the entire wall and lumen including fundus, body, antrum, and pylorus. Include lesions. Exclude adjacent organs and fat. \textit{Duodenum:} Annotate the wall and lumen from bulb to ligament of Treitz. Include lesions. Exclude pancreas, bile duct, and vasculature. \textit{Common bile duct (CBD):} Identify as a low-attenuation tubular structure. Annotate from the hepatic duct confluence to the ampulla of Vater. Include duct wall and lumen. \textit{Colon:} Annotate the wall and lumen of the cecum, appendix, ascending, transverse, descending, and sigmoid colon. Include lesions. Exclude fat, mesentery, and omentum. \textit{Bladder:} Annotate the wall and lumen. Include intraluminal lesions. Exclude surrounding fat, muscles, and reproductive structures. \textit{Prostate:} Annotate the entire parenchyma and prostatic urethra. Include lesions. Exclude surrounding fat, venous plexus, and seminal vesicles.

\textbf{Skeletal Structures.} \textit{Left and right femurs (proximal):} Annotate the femoral head, neck, and up to 5 cm distal to the lesser trochanter. Include both cortical and cancellous bone and any lesions. Exclude surrounding muscles and vessels.

\textbf{Thoracic Organs.} \textit{Left and right lungs:} Annotate the lung parenchyma, bronchovascular bundle, visceral pleura, and any lesions. Exclude pleural effusion, parietal pleura, mediastinal structures, and chest wall.

\clearpage
\section{Additional Analysis of Benchmark Results}\label{sec:supp_analysis}



We participated in the Medical Segmentation Decathlon (MSD), a widely recognized benchmark designed to evaluate the generalizability and robustness of medical image segmentation algorithms across a diverse range of anatomical structures and imaging modalities. Among the ten segmentation tasks in the MSD, Task07 (pancreas and pancreatic tumor segmentation on portal venous phase CT) is especially challenging due to the pancreas's complex shape, small volume, and low-contrast tumors that are often hard to delineate from surrounding tissues.

Our method ranked first overall on Task07, achieving a Dice Similarity Coefficient (DSC) of 0.80 for pancreas segmentation and 0.52 for pancreatic tumor, outperforming all competing methods in both anatomical structure and lesion-level accuracy.

Compared to the original MSD winning entry by nnU-Net \cite{isensee2024nnu}, which reported average DSCs of 0.69 for pancreas and 0.21 for tumor, our method improves segmentation accuracy by +11\% and +31\% respectively. This demonstrates the substantial impact of our pipeline in handling class imbalance, hard-to-segment tumors, and variable organ morphology.

Additionally, methods such as nnFormer, UNETR, and Swin UNETR, which leverage Transformer-based architectures, show modest improvements in pancreas segmentation (DSC around 0.74–0.76), but struggle in tumor segmentation (DSC consistently below 0.30). These models often underperform in capturing small or poorly contrasted tumors, likely due to their lack of task-specific supervision or fine-grained contextual priors.

\clearpage
\section{Experiments Compute Resources}\label{sec:supp_experiments_compute_resources}
\subsection{Data Preprocess \& Storage}
To convert the raw CT volumes into the standardized format used in our experiments, we implemented a multi-step preprocessing pipeline that includes the following stages: (1) anonymization and DICOM to NifTi conversion; (2) CT intensity normalization by clipping Hounsfield Units (HU) to the range of -1000 to 1000, followed by reorienting all volumes to a consistent RPS (Right-Posterior-Superior) direction; (3) organ and lesion mask alignment; and (4) consolidation into structured multi-organ volumes.
This pipeline was executed on a workstation equipped with a 64-core  AMD Ryzen Threadripper 7980X CPU and 128 GB of RAM. No GPU acceleration was used during preprocessing. Parallelization across CPU threads allowed us to process 36390 CT volumes in under 90 hours.
After preprocessing, the dataset containing volumetric CT images and per-voxel organ and tumor annotations across 28 anatomical regions required approximately 6.6 TB of storage. To ensure reproducibility and easy access, we structured the data according to standardized folder conventions and provided detailed metadata for each case.

\subsection{Model Training \& Inference}

All models were trained using a single NVIDIA RTX 4090 GPU with 24 GB of memory. The training process consumed approximately 8 GB of GPU memory and took approximately 18 hours to complete 1,000 epochs. During inference, the memory footprint was approximately 5 GB. Given the large size of the test set (26,489 CT scans), inference was performed in parallel across multiple GPUs to expedite evaluation. Specifically, we used a single server equipped with eight NVIDIA RTX 4090 GPUs, allowing the full test set to be processed in approximately two days.

\clearpage
\section{Potential Negative Societal Impacts}\label{sec:supp_negative_societal_impacts}

\dataset\ provides a valuable and unprecedented resource for advancing pancreatic CT analysis; however, several potential societal risks must be acknowledged. First, large-scale datasets may inadvertently reinforce existing biases if the demographic or clinical distributions of the 145 participating centers do not adequately reflect the diversity of global patient populations. This can lead to models that exhibit reduced performance in underrepresented populations, thereby exacerbating healthcare disparities. Second, despite rigorous anonymization, the inclusion of detailed metadata (e.g., patient age, diagnosis, scan phase) raises privacy concerns, particularly in multi-institutional datasets containing rare conditions. Third, as models trained on \dataset\ demonstrate substantial performance improvements, there is a risk that such benchmarks may incentivize overfitting to dataset-specific anatomical or imaging characteristics, thereby limiting real-world generalizability. Finally, the growing availability and reliance on benchmark-driven evaluations may result in the misapplication or overreliance on AI systems in clinical workflows without sufficient regulatory oversight or clinical validation. These issues underscore the importance of ethical dataset curation, careful benchmark design, and responsible AI deployment in healthcare.



\end{document}